\documentclass[aps,prb,twocolumn,superscriptaddress]{revtex4-1}

\usepackage{hyperref} 
\usepackage{graphicx}
\usepackage{bm}
\usepackage{amsmath}
\usepackage{amssymb}
\usepackage{color,soul}

\DeclareGraphicsExtensions{.png,.jpg,.eps}
\usepackage{xcolor}

\newcommand{\beq}{\begin{eqnarray}}
\newcommand{\eeq}{\end{eqnarray}}
\newcommand*{\citen}[1]{
	\begingroup
	\romannumeral-`\x
	\setcitestyle{numbers}
	\cite{#1}
	\endgroup
}
\begin{document}

\author{Guangze Chen}
\affiliation{Institute for Theoretical Physics, ETH Zurich, 8093 Zurich, Switzerland}
\affiliation{Department of Applied Physics, Aalto University, 02150 Espoo, Finland}

\author{Oded Zilberberg}
\affiliation{Institute for Theoretical Physics, ETH Zurich, 8093 Zurich, Switzerland}

\author{Wei Chen}
\email{Corresponding author: pchenweis@gmail.com}
\affiliation{Institute for Theoretical Physics, ETH Zurich, 8093 Zurich, Switzerland}
\affiliation{National Laboratory of Solid State Microstructures and School of Physics, Nanjing University, Nanjing 210093, China}
\affiliation{College of Science, Nanjing University of Aeronautics and Astronautics, Nanjing 210016, China}

\title{Detection of Fermi Arcs in Weyl Semimetals through Surface Negative Refraction}

\begin{abstract}
One of the main features of Weyl semimetals is the existence of Fermi arc surface states at their surface, which cannot be realized in pure two-dimensional systems in the absence of many-body interactions. Due to the gapless bulk of the semimetal, it is, however, challenging to observe clear signatures from the Fermi
arc surface states. 
Here, we propose to detect such novel surface states via perfect negative refraction that occurs between two adjacent open surfaces with properly orientated Fermi arcs.
Specifically, this phenomenon visibly manifests in non-local transport measurement, where the negative refraction generates a return peak in the real-space conductance. This provides a unique signature of the Fermi arc surface states. We discuss the appearance of this peak both in inversion and time-reversal symmetric Weyl semimetals, where the latter exhibits conductance oscillations due to multiple negative refraction scattering events.
\end{abstract}

\date{\today}

\maketitle

\section{introduction}

In recent years, the classification of topological phases
of matter has been extended from topological insulators
\cite{hasan2010colloquium,qi2011topological} to topological
semimetals~\cite{armitage2018weyl,Fang16cpb}. The latter involves gapless band structures with nontrivial topological properties. Depending on whether the gap closing occurs at isolated
points in the Brillouin zone or along closed loops, they are mainly divided into Weyl/Dirac semimetals~\cite{wan2011topological,murakami2007phase,Burkov11prl,wang2013three,Weng15prx,huang2015weyl}
and nodal-line semimetals~\cite{Burkov11prb}. The unique topological properties of these gapless band structures are extensively explored using a wide variety of platforms, including solid state materials~\cite{wan2011topological,murakami2007phase,Burkov11prl,wang2013three,Weng15prx,
huang2015weyl,lv2015experimental,xu2015discovery,xu2015discovery2,xu2015experimental,
xu2016observation,deng2016experimental,yang2015weyl,huang2016spectroscopic,tamai2016fermi,
jiang2017signature,belopolski2016discovery,lv2015observation,chen2018proposal,chen2019interaction}, but also using photonic~\cite{lu2013weyl,lu2015experimental}, phononic~\cite{xiao2015synthetic,yang2016acoustic}, and electric-circuit~\cite{lee2018topolectrical,luo2018topological,lu19prb} metamaterials.

In Weyl semimetals, the gap closes at so-called Weyl points that
are topologically robust against local perturbations in reciprocal
space~\cite{Balents11phys}, which is beneficial for their
experimental detection~\cite{lv2015experimental,xu2015discovery,xu2015discovery2,xu2015experimental,xu2016observation,deng2016experimental,yang2015weyl,huang2016spectroscopic,tamai2016fermi,jiang2017signature,belopolski2016discovery,lv2015observation}. 
The band topology of Weyl semimetals is encoded in the
monopole charge or Chern number of Berry curvature field carried by each Weyl point.
According to the topological bulk-boundary correspondence of Weyl semimetals, disconnected
Fermi arcs appear in the surface Brillouin zone and span between the Weyl points~\cite{wan2011topological,Yang11prb}.
Such exotic Fermi arcs serve as the fingerprint of Weyl semimetals,
and their experimental identification has attracted great research interest~\cite{lv2015experimental,xu2015discovery,xu2015discovery2,xu2015experimental,xu2016observation,deng2016experimental,yang2015weyl,
huang2016spectroscopic,tamai2016fermi,jiang2017signature,belopolski2016discovery}.

Recent progress has been made on the
observation of Fermi arc states in Weyl and Dirac semimetals by using
angle-resolved photoemission spectroscopy (ARPES)
\cite{lv2015experimental,xu2015discovery,xu2015discovery2,xu2015experimental,
xu2016observation,deng2016experimental,yang2015weyl,huang2016spectroscopic,tamai2016fermi,jiang2017signature,belopolski2016discovery} 
and quantum transport measurement~\cite{Moll16nat,Wang16nc}. 
In these experiments, both bulk and surface states
appear in the measured observables, making it difficult
to explicitly identify the Fermi arcs. Several phenomenon dominated by Fermi arc surfaces states are predicted\cite{bovenzi2017chirality,faraei2019induced,adinehvand2019sound}, yet to be observed. Therefore, there is a need to explore novel and unique transport properties that can facilitate the identification of Fermi
arcs. Moreover, such particular transport signatures open an avenue for their control and manipulation for potential applications ~\cite{chen2019arxiv2}.

\begin{figure}
\center
\includegraphics[width=\linewidth]{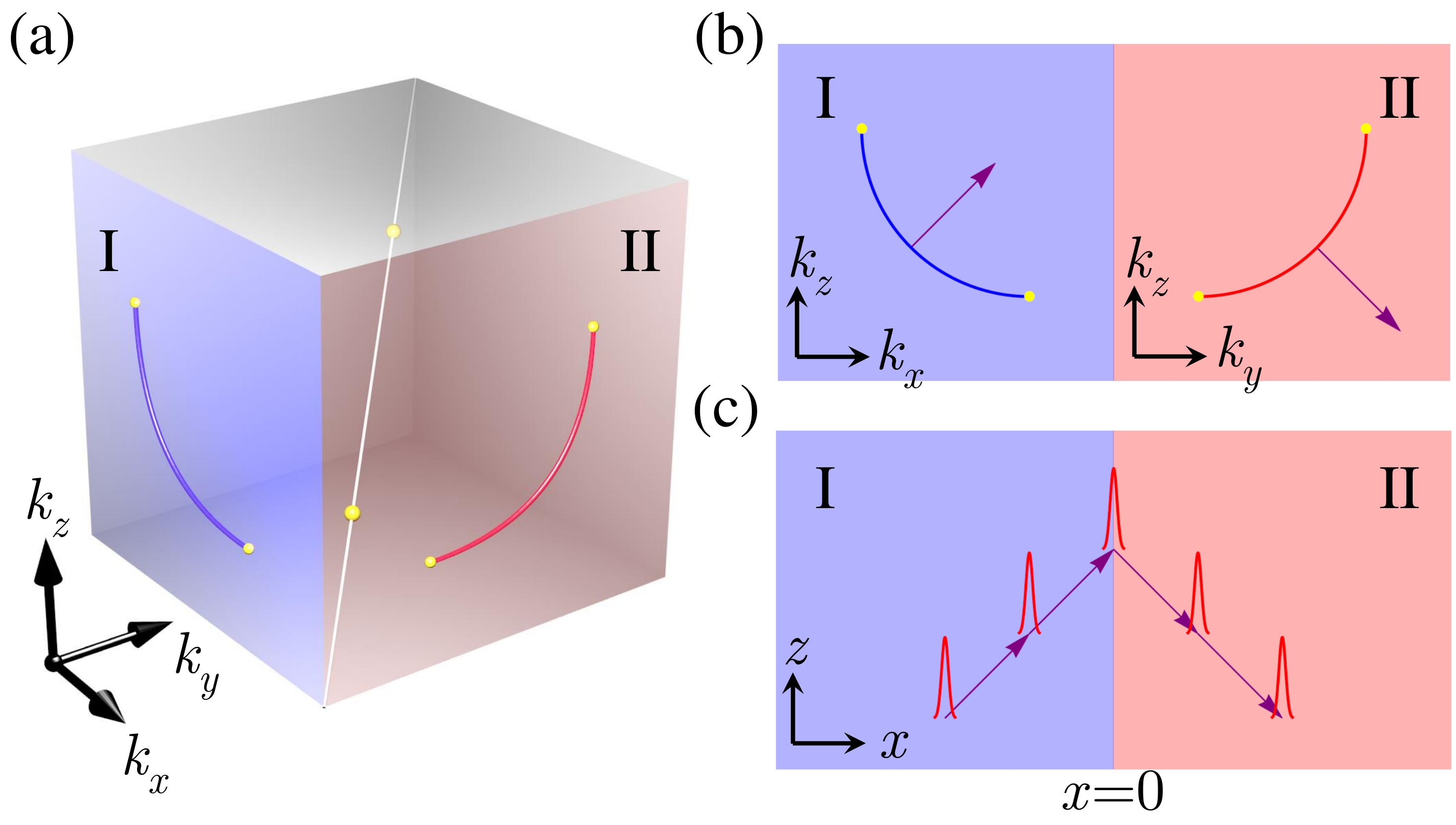}
\caption{ Negative refraction between Fermi arcs at different surfaces of Weyl semimetals. (a) Sketch of a Weyl semimetal with oriented Fermi arcs (red and blue curves). (b) The red and blue surfaces form a junction that can be represented as a 2D scattering problem. The Fermi velocities (purple arrows) have opposite components parallel to the scattering line. (c) Perfect negative refraction of a surface wavepacket due to the tilting of the Fermi arcs.}
\label{fig1}
\end{figure}

Fermi arcs indicate strong anisotropy that breaks rotational symmetry, in contrast to closed Fermi surfaces
in normal metals. As a result, part of the scattering channels at the Fermi energy level
is absent, serving as a source for unique transport properties including negative refraction between different surfaces~\cite{he2018topological}.
In reality, the electronic transport signatures will depend on the material details and their specific termination, both of which affect the Fermi arcs' orientation, dispersion, and length~\cite{lv2015experimental,xu2015discovery,xu2015experimental,xu2015discovery2,xu2016observation,deng2016experimental,MnBi2Te4,wang2019single,soh2019ideal}. 
Notably, however, state-of-the-art fabrication techniques allow for controlled surface shaping on the level of a single layer~\cite{yang2015weyl,xu2016observation,morali2019fermi,yang2019topological}, 
making it possible to explore the broad breadth of surface transport phenomena.

In Ref.\citen{chen2019arxiv2}, it was shown that perfect negative refraction occurs
between two adjacent open surfaces when the respective Fermi arcs are properly orientated. In this work, we show that this scenario manifests for both $\mathcal{P}$- and $\mathcal{T}$-symmetric Weyl semimetals, which generates distinct spatial trajectories for electron propagation.
In particular, we propose to detect the negative refraction via non-local scanning tunneling spectroscopy. The negative refraction manifests as a clear spatially-resolved peak in the non-local conductance.
Adverse effects, such as surface disorder and dispersive corrections to the Fermi arcs do not qualitatively change this transport peak.
Our results offer a decisive signature for the detection of the Fermi arcs and present Weyl semimetals surface transport as a new platform to observe electronic negative refraction~\cite{cheianov2007focusing,cserti2007caustics,beenakker2008colloquium,lee2015observation}. Experimental realization of our proposal is within reach as the surface Fermi arcs orientation can be readily controlled by proper choice the material termination~\cite{yang2015weyl,xu2016observation,morali2019fermi,yang2019topological}.

The paper is organized as follows: in Sec. \ref{arc2},
we show that arbitrary orientations of Fermi arcs
can be described by a rotation transformation of an effective
Hamiltonian. Based on the resulting effective surface
Hamiltonian and using a tunneling approach, we calculate
the non-local conductance between two local terminals
in both inversion ($\mathcal{P}$) and time-reversal ($\mathcal{T}$) symmetric Weyl semimetals
in Sec. \ref{secp} and Sec. \ref{sect}, respectively. This serves as a direct
signature of negative refraction.
Finally, we discuss the experimental
realization of our proposal and draw conclusions in Sec. \ref{secs}.

\section{Oriented Fermi arcs}\label{arc2}

In Weyl semimetals, Fermi arcs appear in
the surface Brillouin zone, connecting
the projection of two bulk Weyl points with opposite monopole charges.
Within the surface Brillouin zone, the orientation of the Fermi arcs depends on
the alignment of the bulk Weyl points relative
to the termination direction of the sample.
Therefore,
by proper cutting of the sample,
different orientations of the Fermi arcs can be obtained.
To describe this orientation dependence, it is convenient
to rotate the effective bulk Hamiltonian of the Weyl semimetal relative
to fixed termination directions\cite{chen2013specular}.

More concretely, we first consider the following minimal model of a
$\mathcal{P}$-symmetric Weyl semimetal
\begin{eqnarray} \label{p}
H(\bm{k})=\hbar v(k_x\sigma_x+k_y\sigma_y)+M(k_0^2-\bm{k}^2)\sigma_z,
\end{eqnarray}
where $v$, $k_0$ and $M>0$ are parameters,
$\bm{k}=(k_x, k_y, k_z)$ is the wave vector, and $\sigma_{x,y,z}$ are Pauli
matrices acting on the 2-band pseudospin space.
By diagonalizing the Hamiltonian, one can find two Weyl points located at
$\pm\bm{k}_0=(0,0,\pm k_0)$. We calculate the topologically-protected
surface states at an open surface in the $-y$
direction (surface I in Fig. \ref{fig1}). They are confined by $k_x^2+k_z^2<k_0^2$
and described by the effective Hamiltonian (see Appendix \ref{analytical_arc})
\beq \label{s1}
H^0_{\text{I}}(k_x,k_z)=\hbar vk_x.
\eeq
Similarly, the surface states on the open surface in the
$x$ direction (surface II in Fig.\ref{fig1}) are described by
\beq\label{s2}
H^0_{\text{II}}(k_y,k_z)=\hbar vk_y.
\eeq
On both surfaces, the states are parallel to the $z$-direction. Therefore, Fermi
arcs states at  a chemical potential within the bulk gap (henceforth
taken at $E=0$) are also parallel to the $z$-direction. Correspondingly,
due to the chirality of the surface states, electrons are fully
transmitted without backscattering at a junction between the
surfaces I and II, see Fig. \ref{fig1}(b).

Next, we perform a rotational transformation to the effective bulk
Hamiltonian Eq. \eqref{p}. In this way, we retain the same
open boundary conditions and describe generally-orientated
Fermi arcs.
A rotation about the axis $k_x=k_y,k_z=0$
by an angle $\varphi$ is defined by
$
H'(\bm{k})= H(U^{-1}\bm{k})
$
with the rotation operator
\beq \label{eq4}
U(\varphi)=
\small{\left(\begin{array}{ccc}\cos^2\frac{\varphi}{2}
&\sin^2\frac{\varphi}{2}&-\frac{\sin\varphi}{\sqrt{2}}
\\\sin^2\frac{\varphi}{2}&\cos^2\frac{\varphi}{2}
&\frac{\sin\varphi}{\sqrt{2}}\\\frac{\sin\varphi}{\sqrt{2}}
&-\frac{\sin\varphi}{\sqrt{2}}&\cos\varphi\end{array}\right)}.
\eeq
As a result, the bulk Weyl points are located at $U\bm{k}_0=\pm k_0(-\frac{\sin\varphi}{\sqrt{2}},\frac{\sin\varphi}{\sqrt{2}},\cos\varphi)$ and the states on surface I can be
described by the effective Hamiltonian
\beq
H_{\text{I}}(k_x, k_z)=\hbar v'(\cos\theta k_x+\sin\theta k_z),
\eeq
where $v'$ is the
renormalized velocity
and $\theta=\tan^{-1}(\tan\varphi/\sqrt{2})$.
The Fermi arc defined by $H_{\text{I}}=0$ is
\beq
\cos\theta k_x+\sin\theta k_z=0,
\eeq
and stretches between $\pm k_0(-\frac{\sin\varphi}{\sqrt{2}},\cos\varphi)$. Note that our approach of rotating the effective bulk model and calculating the resulting surface dispersion is verified using microscopic lattice model simulations, see Appendix \ref{arc}. Similarly, on surface II
\beq
H_{\text{II}}(k_y, k_z)=\hbar v'(\cos\theta k_y-\sin\theta k_z),
\eeq
and the Fermi arc is defined by
\beq
\cos\theta k_y-\sin\theta k_z=0,
\eeq
and stretches between $\pm k_0(\frac{\sin\varphi}{\sqrt{2}},\cos\varphi)$.
Note that the two Fermi arcs have different orientations; see Fig. \ref{fig1}.
For a finite $\theta$, electrons incident on surface I
can only transfer through the interface due to the lack of
backscattering channels. At the same time, because the Fermi
arcs on the two surfaces tilt in opposite directions,
the velocity in the $z$-direction
is inverted, leading to negative refraction as shown
in Figs. \ref{fig1}(b) and (c).

In the following, we introduce a general dispersion term
to the surface Hamiltonian
\begin{equation}\label{HS}
\begin{split}
H'_{\text{I}}(k_x,k_z)&=H_{\text{I}}+\varepsilon_x,\\
H'_{\text{II}}(k_y,k_z)&=H_{\text{II}}-\varepsilon_y
\end{split}
\end{equation}
with a parabolic dispersion $\varepsilon_{x,y}=d[k_0^2(1-\sin^2\varphi/2)-k_{x,y}^2-k_z^2]$.
By tuning the dispersion strength $d$, the Fermi
arcs become curved; see Figs. \ref{fig1} and \ref{fig2}(b). Such curving captures
the situation in real materials~\cite{lv2015experimental,xu2015discovery,xu2015discovery2,xu2015experimental,
xu2016observation,deng2016experimental,yang2015weyl,
huang2016spectroscopic,tamai2016fermi,jiang2017signature,belopolski2016discovery,lv2015observation}.
Moreover, the velocities of the surface states
are also modified. In our following calculation,
we assume that the dispersion
does not invert the velocity in
the $x$- ($y$-) direction on surface I (II).
Note that the description of
generally orientated
Fermi arcs by rotation of the effective model
works for both $\mathcal{P}$- and $\mathcal{T}$-symmetric Weyl
semimetals. This approach is verified by numerical simulations
of corresponding lattice models (see Appendix \ref{arc}).

\section{Negative refraction in $\mathcal{P}$-symmetric Weyl semimetals}\label{secp}
\begin{figure}[htbp]
\center
\includegraphics[width=\linewidth]{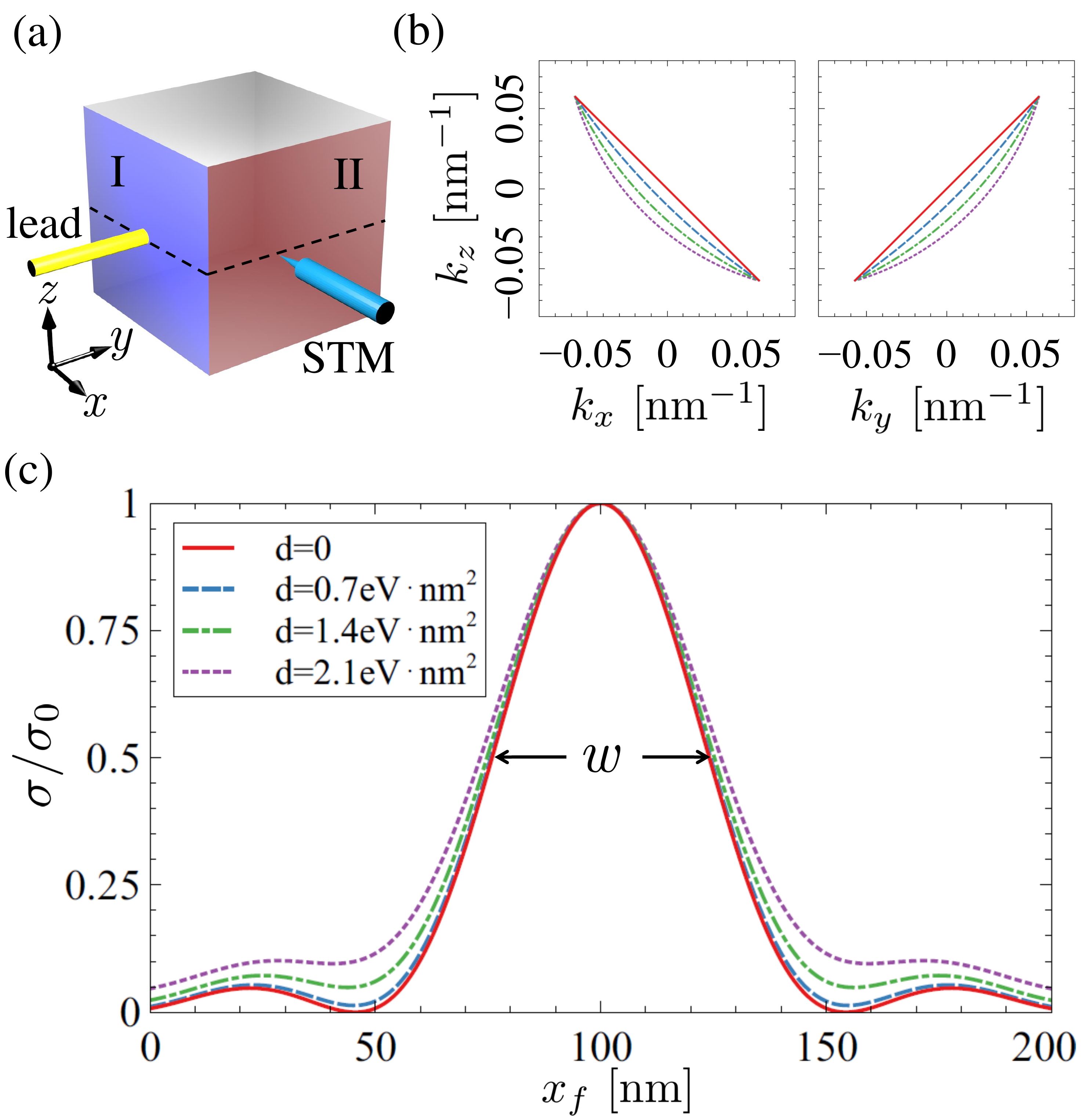}
\caption{ Non-local conductance between surfaces of Weyl semimetals. (a) Sketch of the setup. (b)
Fermi arcs with different curvature controlled
by $d$ [cf. Eq. \eqref{HS}], labelled by the legend in (c).
(c) Non-local conductance $\sigma(\varepsilon=0)$
for Fermi
arcs with different curvature [cf. Eq. \eqref{cod}], with parameters:
$\theta=\frac{\pi}{4}$, $k_0=0.1$nm$^{-1}$,
$v'=10^6$m/s, and $x_i=100$nm. The peak structure
indicates the existence of negative refraction. The peak width $w$ is comparable with $\pi/k_0$. }
\label{fig2}
\end{figure}

Next, we investigate nonlocal electron transport through the
surface states, see the corresponding
two-terminal setup in Fig. \ref{fig2}(a).
For convenience, we unfold the two open surfaces
to the $x-z$ plane with the boundary located at $x=0$
[Fig. \ref{fig1}(c)], which can be achieved by
the replacement $H'_{\text{II}}(k_y\rightarrow k_x,k_z)$
in Eq. \eqref{HS}.
An electron wave packet is injected from the local lead at
$\bm{r}_i=(-x_i,0)$ on surface I,
then transmitted to surface II via negative
refraction,
and finally reaches the tip of the scanning tunneling microscope (STM)
at $\bm{r}_f=(x_f,0)$.
The wave packet propagates along a spatially-localized
trajectory [cf. Fig. \ref{fig1}(c)]. This behavior can be
revealed by the appearance of a peak structure in the spatially resolved
non-local conductance as a function of $x_f$ [calculated below]; see Fig. \ref{fig2}(c).
Crucially, this signature is unique to the negative refraction through the Fermi
arc surface states. For normal metal states,
the conductance decays with $x_f$, as the wave packet
expands in the $z$-direction.

In the following,
we calculate the non-local conductance
using the surface Hamiltonian \eqref{HS} and the
Green's function method.
The Fermi energy is set to zero for simplicity,
so that bulk electrons do not contribute to the
conductivity. The finite density of the bulk states can solely
lead to leakage of electrons, which will not
change our main results. The coupling between
the terminals and the surface states is described by a tunneling Hamiltonian as
\beq\label{T}
H_T
&=&\sum_{p,\alpha=i,f}T_\alpha d^\dag_{p,\alpha}\Psi(\bm{r}_\alpha)+\text{H.c.}
\eeq
where $T_{\alpha}$ is the tunneling strength between the
system and the $\alpha$ terminal, $d_{p,\alpha}$ is the Fermi operator
in the $\alpha$ terminal with momentum $p$ and $\Psi(\bm{r})$ is the field operator of the surface states at position $\bm{r}$, with $\bm{r}_\alpha$ corresponding to each terminal location.

The non-local conductance (including spin degeneracy) between
local electrode and the STM tip is given by \cite{datta1997electronic}
\beq\label{cod}
\sigma(\varepsilon)=\frac{2e^2}{h}\text{Tr}[\Gamma_iG^R\Gamma_fG^A].
\eeq
The full retarded ($R$) and advanced ($A$) Green's function
$G^{R,A}$ and the linewidth functions $\Gamma_\alpha$
are [see Appendix \ref{gf} for details]
\beq \label{G}
\resizebox{.85\hsize}{!}{$\displaystyle
G^{R,A}_\varepsilon(\bm{r}_f,\bm{r}_i)=(1+R_i)^{-1} g_\varepsilon^{R,A}(\bm{r}_f,\bm{r}_i)(1+R_f)^{-1},$}
\eeq
\beq \label{L}
\resizebox{.85\hsize}{!}{$\displaystyle
\Gamma_{\alpha}(\bm{r}_1,\bm{r}_2,\varepsilon)
=2\pi\rho_{\alpha}(\varepsilon)|T_{\alpha}|^2\delta(\bm{r}_1-\bm{r}_{\alpha})\delta(\bm{r}_2-\bm{r}_{\alpha}),$}
\eeq
where the function $R_{\alpha}(\varepsilon)=\pi^2\rho_{0}(\varepsilon)\rho_{\alpha}(\varepsilon)|T_{\alpha}|^2$
with $\rho_0(\varepsilon)=k_0/(2\pi^2\hbar v)$ the density of states (DOS) of Fermi arc
surface states per unit area and $\rho_\alpha(\varepsilon)$ the DOS of the terminal $\alpha$ at energy $\varepsilon$.
The bare Green's function are [cf.~Eq.~\eqref{eqB2}]
\beq\label{g}
g^{R}_\varepsilon(\bm{r}_f,\bm{r}_i)=[g_\varepsilon^A(\bm{r}_i,\bm{r}_f)]^*=-2\pi i\rho_0(\varepsilon) f_\varepsilon(\bm{r}_f,\bm{r}_i),
\eeq
with
\beq\label{f}
\resizebox{.85\hsize}{!}{$\displaystyle
f_\varepsilon(\bm{r}_f,\bm{r}_i)=\int_{-k_0\cos\varphi}^{k_0\cos\varphi} dk_z\frac{ e^{i(k_{x2}x_f-k_xx_i)}}{2k_0\cos\varphi}e^{ik_z(z_f-z_i)}.$}
\eeq
Here,  $k_x$ and $k_{x2}$ are solved by $H'_{\text{I}}(k_x,k_z)=\varepsilon$
and $H'_{\text{II}}(k_{x2},k_z)=\varepsilon$, respectively. The interval of integration covers the Fermi arc region,
and the $k_z$ dependence of the velocity in the $x$-direction is ignored.

\begin{figure}[htbp]
\center
\includegraphics[width=\linewidth]{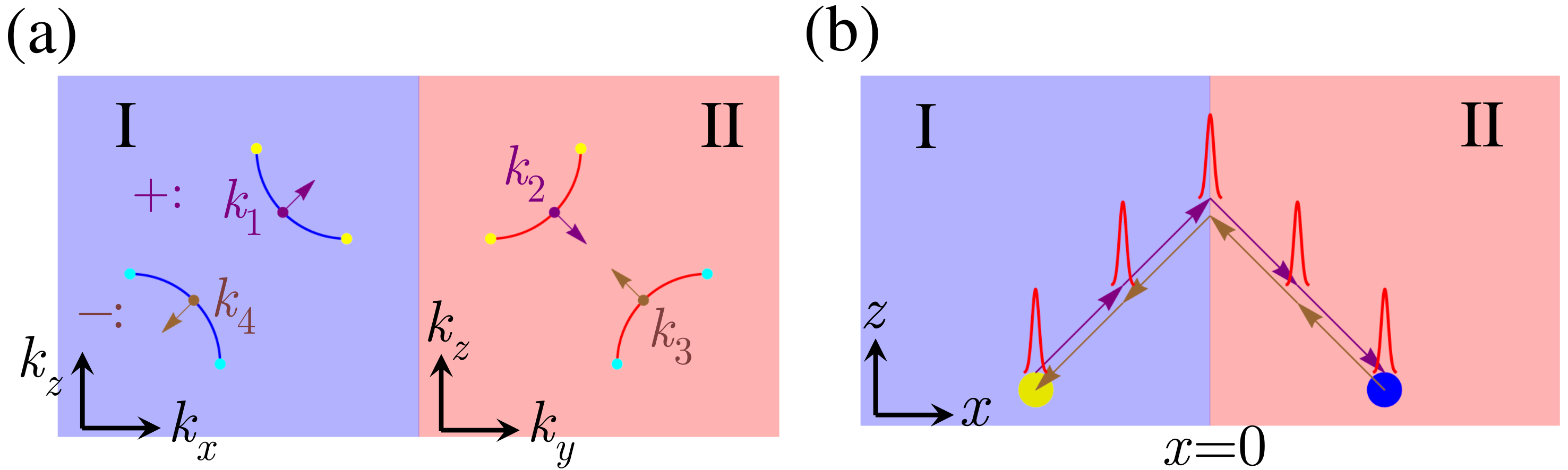}
\caption{ Negative refraction between multiple Fermi arcs at different surfaces of Weyl semimetals. (a): Effective surface model for a $\mathcal{T}$
-symmetric Weyl semimetal. There are two branches of Fermi arcs with opposite chirality (labeled as "$+$" and "$-$", respectively). (b):
Fabry-P\'{e}rot interference led by backscattering at the terminals (blue and gold disks).}
\label{fig3}
\end{figure}

Performing integration in Eq. \eqref{cod} yields
\beq\label{codp}
\sigma(\varepsilon)&=&\sigma_0(\varepsilon)\left|f_\varepsilon(\bm{r}_f,\bm{r}_i)\right|^2,\\
\sigma_0(\varepsilon)&=&\frac{32e^2}{h}\frac{R_i}{(1+R_i)^2}\frac{R_f}{(1+R_f)^2},
\eeq
where $\sigma_0$ takes the maximum value
$2e^2/h$ when $R_i=R_f=1$.
The dependence of $\sigma(\varepsilon)$ on $x_f$
comes from the factor $\left|f_\varepsilon(\bm{r}_f,\bm{r}_i)\right|^2$,
which has a peak due to negative refraction; see Fig. \ref{fig2}(c).
In particular, when $\varepsilon=0$, $k_{x2}(\varepsilon,k_z)=-k_x(\varepsilon,k_z)$,
one can see
from Eq. \eqref{f} that the peak of the function
$\left|f_\varepsilon(\bm{r}_f,\bm{r}_i)\right|^2$
is centered around $x_f=x_i$
on the $x$ axis.
The peak structure in the non-local conductance
stems from the wave packet trajectory of negative
refraction in Fig. \ref{fig1}(c). The width of the peak $w$,
corresponding to the scale of the wave packet,
is comparable with $\pi/k_0$, which can be seen from Eq.~\eqref{f}. Specifically, in the case of straight Fermi arcs and $\varepsilon=0$, we will have $k_{x2}=-k_x=k_z$. Performing the integration in Eq.~\eqref{f} yields $f_{\varepsilon=0}(\bm{r}_f,\bm{r}_i)=\sin(k_0(x_i+x_f)\cos\varphi)/(k_0(x_i+x_f)\cos\varphi)$, thus the peak width for $|f_{\varepsilon=0}(\bm{r}_f,\bm{r}_i)|^2$ is comparable with $\pi/k_0$.
For curved Fermi arcs with dispersion ($d\neq0$),
the wave packet spreads during its propagation,
so that the peak of conductance is broadened as well;
see Fig. \ref{fig2}(c). Therefore, the peak width $w$ provides useful information about the length of the Fermi arcs. The existence of the peak structure is also confirmed numerically in Fig. \ref{numerics}(a).

\section{Negative refraction in $\mathcal{T}$-symmetric Weyl semimetals}\label{sect}

In reality, there are only few material candidates for Weyl semimetals
with only two Weyl points~\cite{MnBi2Te4,wang2019single,soh2019ideal}.
Hence, we investigate negative refraction between the surface states of
$\mathcal{T}$-symmetric Weyl semimetals, which are more abundant~\cite{lv2015experimental,xu2015discovery,
xu2015experimental,xu2016observation,xu2015discovery2}.
Specifically, we study a semimetal with four Weyl points.
Our results can be readily extended to
the situation with more Weyl points.

Consider a $\mathcal{T}$-symmetric Weyl semimetal
with two pairs of Weyl points.
Correspondingly, there are two Fermi arc segments on
each open surface, which are the time-reversal counterpart to
each other; see Fig. \ref{fig3}(a). The existence of two branches of
surface states with opposite chirality enables backscattering between them.
For simplicity, we restrict our discussion to the case
that two Fermi arcs on the same surface
do not overlap when projecting to the $k_z$ axis.
It means that no backscattering occurs for conserved $k_z$,
so that perfect negative refraction occurs at the interface
between surfaces I and II \cite{he2018topological}.
However, backscattering takes place at the local terminals,
leading to Fabry-P\'{e}rot interference [Fig. \ref{fig3}(b)]
and additional oscillation
of the non-local conductance on top of the peak structure in real space.

More concretely, the two adjacent open surfaces I and II contains
two Fermi arcs each, as shown in Fig. \ref{fig3}(a).
We describe the branch ``$+$" by
\beq \label{effective_surface_TR}
H_{+}(k_x,k_z)=\left\{\begin{array}{cc}H'_{\text{I}}
(k_x-k_{x_0},k_z-k_{z_0})&x<0\\H'_{\text{II}}
(k_x+k_{x_0},k_z-k_{z_0})&x>0\end{array}\right.,
\eeq
which is similar to Eq. \eqref{HS} except for a
shift of the Fermi arcs in the surface Brillouin zone.
The time-reversal counterpart, branch ``$-$" is
described by $H_{-}(k_x,k_z)=H_{+}(-k_x,-k_z)$.

\begin{figure}[htbp]
\center
\includegraphics[width=\linewidth]{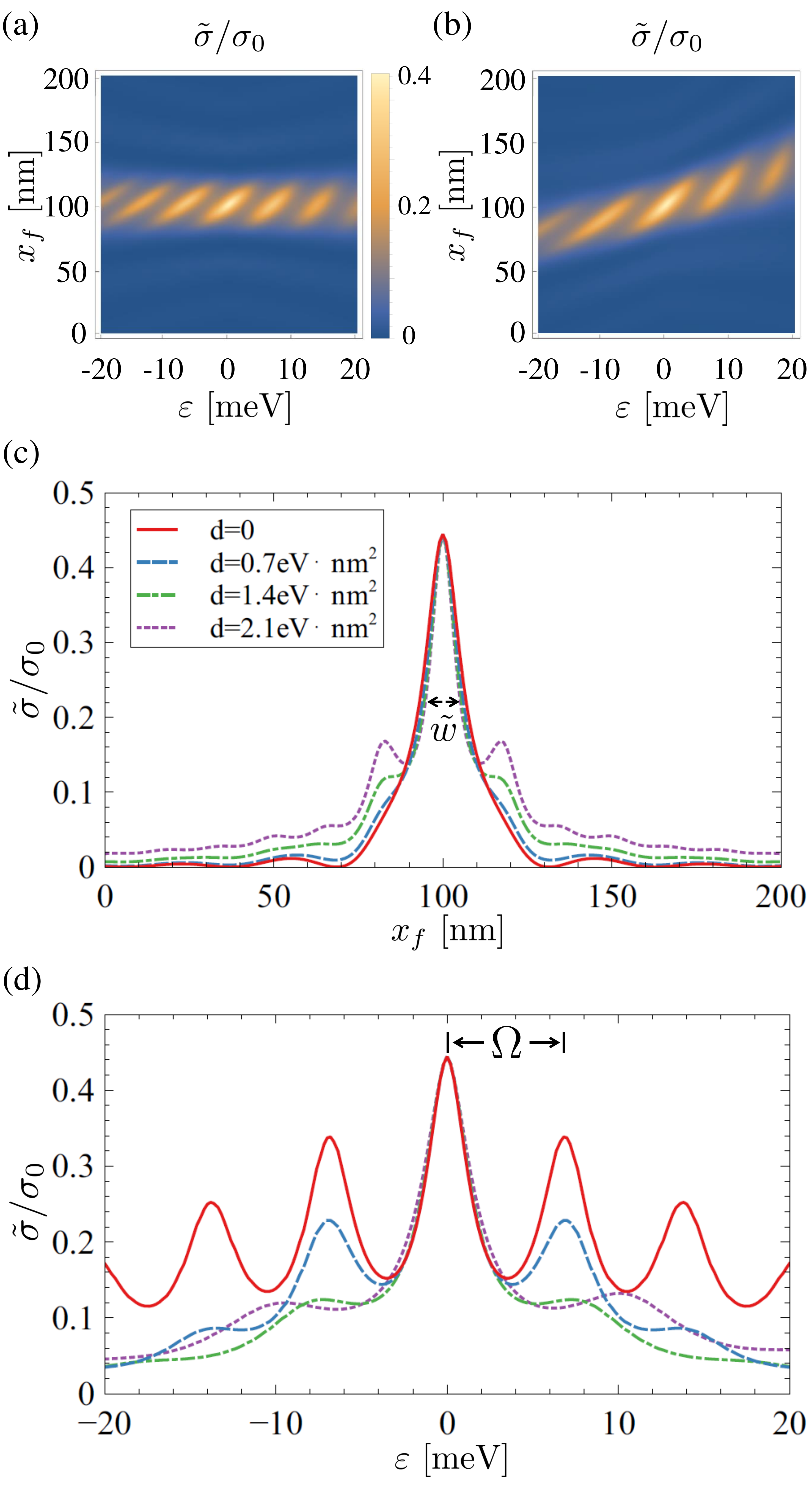}
\caption{Non-local conductance $\tilde{\sigma}(\varepsilon)$ in a
$\mathcal{T}$-symmetric Weyl semimetal [cf. Eq. \eqref{effective_surface_TR}] with (a) straight ($d=0$) and (b) curved Fermi arcs ($d=0.7$eV$\cdot$nm$^2$). The dependence of the conductance on (c) $x_f$ for fixed $\varepsilon=0$ and (d) $\varepsilon$ with $x_f=x_i$ for Fermi arcs with different curvature [labelled by the legend in (c)]. All parameters are $R_i=R_f=1$, $k_0=0.2$nm$^{-1}$, $k_{x_0}=k_{z_0}=k_0/\sqrt{2}$ and other parameters the same as those in Fig.~\ref{fig2}.}
\label{TR_conductance_FP}
\end{figure}

Similar to the $\mathcal{P}$-symmetric case,
we first solve the Green's function
for the surface states, yielding
\beq
\tilde{g}_{\varepsilon}^{R}(\bm{r}_f,\bm{r}_i)=\tilde{g}_{\varepsilon}^{R}(\bm{r}_i,\bm{r}_f)=-\pi i\rho'_0(\varepsilon)f'_\varepsilon(\bm{r}_f,\bm{r}_i).
\eeq
with
\beq\label{f'}
f'_\varepsilon(\bm{r}_f,\bm{r}_i)=\int_{k_{1}^+}^{k_{2}^+}dk_z \frac{e^{i(k'_{x2}x_f-k'_xx_i)}}{k_{2}^+-k_{1}^+}e^{ik_z(z_f-z_i)},
\eeq
where $\rho'_0(\varepsilon)$ is the density of surface states per unit area,
and $k_{1}^+$ and $k_{2}^+$ are the $k_z$ component of the terminations
of the Fermi arcs in the branch ``+". $k'_x$ and $k'_{x2}$ are solved by $H'_{\text{I}}(k'_x-k_{x_0},k_z-k_{z_0})=\varepsilon$
and $H'_{\text{II}}(k'_{x2}+k_{x_0},k_z-k_{z_0})=\varepsilon$, respectively.
We describe the coupling to the terminals by the same tunneling Hamiltonian
\eqref{T}, which leads to
the same self-energy in Eq. \eqref{self-energy_inversion}.
The full Green's function, however, takes a
different form due to the backscattering at the terminals,
\beq
\resizebox{.85\hsize}{!}{$\displaystyle
\tilde{G}_{\varepsilon}^{R}(\bm{r}_f,\bm{r}_i)
=\frac{\tilde{g}_\varepsilon^R(\bm{r}_f,\bm{r}_i)}{(1+R_i)(1+R_f)-R_iR_ff'^2_\varepsilon(\bm{r}_f,\bm{r}_i)}.$}
\eeq
The resulting non-local conductance calculated by Eq. (\ref{cod}) is
\beq \label{codt}
\resizebox{.85\hsize}{!}{$\displaystyle  \tilde{\sigma}(\varepsilon)=\frac{\sigma_0}{4}\left|\frac{f'_\varepsilon(\bm{r}_f,\bm{r}_i)}{1-R_iR_ff'^2_\varepsilon(\bm{r}_f,\bm{r}_i)/[(1+R_i)(1+R_f)]}\right|^2,$}
\eeq
which mainly differs from the $\mathcal{P}$-symmetric Weyl semimetal [cf. Eq. \eqref{codp}]
by the additional
term in the denominator due to
the multiple scattering in Fig. \ref{fig3}(b).
In the weak tunneling limit $R_{i,f}\ll1$,
the effect due to multiple scattering is negligible
and the conductance
$
\tilde{\sigma}(\varepsilon)\approx\frac{\sigma_0}{4}|f'_\varepsilon(\bm{r}_f,\bm{r}_i)|^2.
$
More generally, the conductance as a function
of energy and $x_f$ is plotted in Figs. \ref{TR_conductance_FP}(a) and \ref{TR_conductance_FP}(b).
The non-local conductance displays additional Fabry-P\'{e}rot oscillations
induced by multiple scattering on top of the peak structure in real space, resulting in the appearance of side peaks for large dispersion $d$; see Fig. \ref{TR_conductance_FP}(c), which is verified by numerical simulations using a lattice model in Fig. \ref{numerics}(b). The width of the main peak $\tilde{w}$ is comparable to $\pi/(k_{2}^+-k_1^+)$ for the same reason as in the $\mathcal{P}$-symmetric case.
The Fabry-P\'{e}rot interference also results in oscillation of conductance with energy when $d$ is small [Figs.~ \ref{TR_conductance_FP}(a) and \ref{TR_conductance_FP}(d)], which is due to the dependence of the Fermi momenta on energy. Specifically, assume that, for the branch "+", the Fermi velocity along the x-direction [cf. Fig.~\ref{fig3}], $v_{x,+}$, is independent of $k_z$. This implies that when the energy increases by $\Delta\varepsilon$ the momenta $k_x'$ and $k_{x2}'$ will increase by $\Delta\varepsilon/v_{x,+}$. Therefore, the function $f'_\varepsilon(\bm{r}_f,\bm{r}_i)$ [cf. Eq.~\eqref{f'}] gains an additional phase factor of $\Delta\varepsilon(x_f-x_i)/v_{x,+}$. Due to the factor $f'_\varepsilon(\bm{r}_f,\bm{r}_i)^2$ in the denominator of Eq.~\eqref{codt}, the oscillation period with $\varepsilon$, $\Omega$, is comparable with $\pi v_{x,+}/(x_f-x_i)$.
Note that, in the case of large $d$, the center of the resonant peak in real space moves
with $\varepsilon$ [Fig.~\ref{TR_conductance_FP}(b)], and oscillation
with $\varepsilon$ cannot be seen due to the rapid decrease
of the conductance at $\varepsilon\neq0$ [Fig. \ref{TR_conductance_FP}(d)].

\section{Discussion and conclusion}\label{secs}

\begin{figure}
\center
\includegraphics[width=\linewidth]{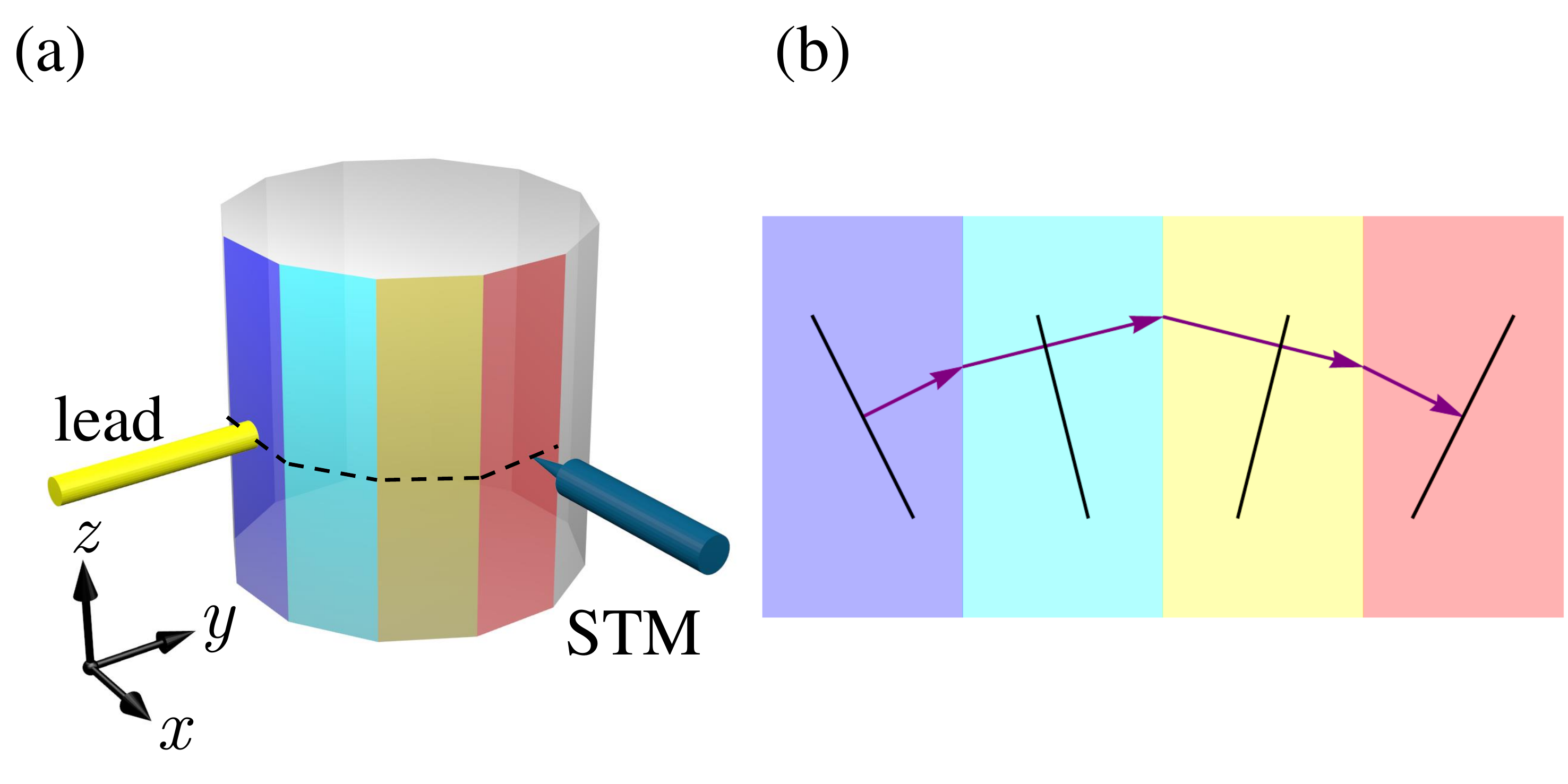}
\caption{(a) Sketch of the setup of nonlocal conductance measurement in the case of a polyhedral nanowire. (b) Tilted Fermi arcs (black lines) result in spatially localized trajectory (purple arrows) where negative refraction takes place on one of the interfaces. Such localized trajectory leads to peak structure in nonlocal conductance similar to Figs.\ref{fig2} and \ref{TR_conductance_FP}.}
\label{different_geometry}
\end{figure}

So far, we have analyzed negative refraction based on the minimal model of
$\mathcal{P}$- and $\mathcal{T}$-symmetric Weyl semimetals.
Several important issues related to the experimental
implementation of our proposal are discussed in the following:

(i) Our scheme applies also to polyhedral nanowires with $N$ surfaces. When the Weyl nodes are aligned in a direction deviating from the central axis, the Fermi arcs on the surfaces become tilted and many refraction processes take place at the boundary of the facets, as shown in Fig. \ref{different_geometry}. While many of the refraction processes are normal refraction, one of them is negative refraction which leads to a spatially localized trajectory similar to Fig.\ref{fig1}(c). The scheme should also hold at $N\to\infty$, when polyhedron becomes a cylinder.

(ii) For Weyl semimetals with more Weyl points and Fermi arcs than
those obtained within the minimal model,
as in most materials\cite{lv2015experimental,
xu2015discovery,xu2015discovery2,
xu2015experimental,xu2016observation,deng2016experimental,yang2015weyl,
huang2016spectroscopic,tamai2016fermi,jiang2017signature,
belopolski2016discovery,lv2015observation}, our main results still hold
as long as the overlap between the projections of different incident and reflection
channels with conserved momentum $k_z$ is negligibly small.
In this case, due to the different orientations of Fermi arcs and the
corresponding trajectories of negative refraction, a multiple peak structure 
in the nonlocal conductance may appear in
the same transport scheme in Fig. \ref{fig2}(a).
The negative refraction will get suppressed if the overlap between
the projections of the incident and reflection Fermi arcs is large
due to the enhanced backscattering.

(iii) We considered Fermi arcs with a regular shape,
such that electrons propagate on the surfaces towards certain directions, which
is the main difference between Fermi arc states and normal metal states.
For Weyl semimetals with long and winding Fermi arcs, surface transport will occur in different directions similar to normal metals, and negative refraction cannot be observed.

(iv) In real materials, the Fermi energy usually deviates from the Weyl points,
resulting in a finite density of bulk states. Our result is not sensitive to
such a deviation because the nonlocal transport occurs on the surface of the sample.
The bulk states only lead to certain leakage of the injected electrons,
and these leaked electrons do not follow the trajectory of negative refraction. As a result, their propagation does not have a peak structure in real space and they contribute only a small background to the conductance peak in the nonlocal transport. Such a small background will not change the qualitative results.

(v) By using Eqs.\eqref{s1} and \eqref{s2} as effective descriptions of the Fermi arc surface states we ignore the penetration of the surface states into the bulk. This is because in most intervals between the Weyl nodes the surface states are well-localized on the surface. Only in the vicinity of Weyl points, will the surface states possess a long penetration into the bulk. These states have no much difference from the bulk states and will not kill the signature of negative refraction as discussed in point (iv). Another effect of such penetration is that it effectively reduces the available transport channels on the surface or equivalently, the length of the Fermi arcs, which also does not change the main results. 

(vi) Finally, surface imperfections such as dangling bonds may exist,
which can be treated as disorder.
In $\mathcal{P}$-symmetric Weyl semimetals,
such surface disorder should have little effect on
the negative refraction and the conductance peak remains stable.
This is because the surface states are unidirectional
and are thus immune to backscattering.
However, in $\mathcal{T}$-symmetric Weyl semimetals
surface disorder will lead to backscattering
between the time-reversal counterpart of the
Fermi arcs with opposite chirality,
which reduces the negative refraction efficiency as well
as the peak structure of the nonlocal conductance.

In summary, we have shown that perfect negative refraction, which can be realized on two adjacent surfaces of Weyl semimetals with properly oriented Fermi arcs, leads to distinct spatial trajectories
for electron propagation.
The space resolved peak structure of the nonlocal conductance 
which indicates the trajectory of negative refraction
can serve as unique evidence of the Fermi arc states.
Recent progress on Weyl semimetals with a single pair of
Weyl nodes in MnBi$_2$Te$_4$\cite{MnBi2Te4}
and EuCd$_2$As$_2$\cite{wang2019single,soh2019ideal}
paves the way to the realization of our proposal.
Furthermore, the manipulation of the negative refraction process offers potential applications of a Weyl semimetal nanowire as a field-effect transistor\cite{chen2019arxiv2}.
Our work opens a new platform
to study negative refraction in electronics \cite{cheianov2007focusing,cserti2007caustics,beenakker2008colloquium,lee2015observation}.
Compared with the existing physical systems, the negative refraction
in Fermi arc states exhibit an unambiguous signature for its detection.

\begin{acknowledgments}
We acknowledge financial support from the Swiss National Science
Foundation through the Division II. We would like to thank Jose Lado for helpful discussions.
\end{acknowledgments}

\begin{appendix}

\section{Derivation of Fermi arc states}\label{analytical_arc}
We derive the Fermi arc surface state at an open surface in the $-y$ direction Eq.\eqref{s1} in the Weyl semimetal Eq.\eqref{p}. The surface state at the open surface in the $x$ direction can be obtained similarly. To calculate the surface state we make the substitution $k_y\to -i\partial_y$ to the Hamiltonian Eq.\eqref{p} since the existence of the open surface breaks translational symmetry in $y$ direction. Thus the surface state $\psi(k_x,y,k_z)$ satisfies the following equation
\beq \label{surface_eq}
(H(k_x,-i\partial_y,k_z)-E)\psi(k_x,y,k_z)=0,
\eeq
with boundary conditions 
\beq \label{eq_A2}
\psi(k_x,y=0,k_z)=\psi(k_x,y=+\infty,k_z)=0.
\eeq
Expanding $\psi(k_x,y,k_z)=\sum_\lambda a_\lambda\psi_\lambda$ on the basis 
\beq
\psi_\lambda(k_x,y,k_z)=e^{ik_xx+ik_zz}e^{\lambda y}\left(\begin{array}{c}a\\b\end{array}\right),
\eeq
where $a$ and $b$ are the pseudo-spin components of $\psi(k_x,y,k_z)$, and substituting into Eq.\eqref{surface_eq}, we have $a_\lambda\neq0$ only for
\beq \label{eq_A4}
(H(k_x,-i\lambda,k_z)-E)\left(\begin{array}{c}a\\b\end{array}\right)=0.
\eeq
Eq.\eqref{eq_A4} implies that 
\beq
H^2=\hbar^2v^2(k_x^2-\lambda^2)+M^2(\lambda^2-F)^2=E^2
\eeq
where $F=k_x^2+k_z^2-k_0^2$, yielding two possible solutions of $\lambda^2$:
\beq \label{eq_A6}
\lambda^2=F+\frac{\hbar^2v^2\pm\sqrt{4M^2F_2+\hbar^4v^4}}{2M^2}
\eeq
with $F_2=E^2+\hbar^2v^2(F-k_x^2)$. Among all possible linear combinations of $\psi_\lambda$ with $\lambda$ satisfying Eq.\eqref{eq_A6}, the only one that satisfies the boundary conditions Eq.\eqref{eq_A2} is
\beq \label{eq_A7}
\psi(k_x,y,k_z)=e^{ik_xx+ik_zz}(e^{\lambda_1 y}-e^{\lambda_2 y})\left(\begin{array}{c}a\\b\end{array}\right)
\eeq
with $\lambda_2<\lambda_1<0$. To get the dispersion of the eigenstate \ref{eq_A7}, note that:
\beq \label{eq_A8}
-\frac{a}{b}=\frac{\hbar v(k_x-\lambda_1)}{M(\lambda_1^2-F)-E}=\frac{\hbar v(k_x-\lambda_2)}{M(\lambda_2^2-F)-E}.
\eeq 
The self-consistent solution to Eqs.\eqref{eq_A6} and \eqref{eq_A8} is 
\beq
E=\text{sgn}(M)\hbar v k_x,
\eeq
with sgn being the sign function.

\section{Numerical calculation of Fermi arcs}\label{arc}
In this Appendix, we verify numerically that for the $\mathcal{P}$- and $\mathcal{T}$- symmetric Weyl semimetals [cf.~Eq.~\eqref{p}], rotation of the effective bulk
model leads to the oriented Fermi arcs on open surfaces.

For the $\mathcal{P}$-symmetric Weyl semimetal we adopt the the effective model $H'(\bm{k})$ in the article. For the $\mathcal{T}$-symmetric Weyl semimetal,
we start with a minimal model
\beq
H_{\text{TR}}(\bm{k})\!=\!M\left(k_1^2-k_x^2\right)\sigma_x\!+\!\hbar vk_y\sigma_y\!+\!M\left(k_0^2-k_y^2-k_z^2\right)\sigma_z,\nonumber\\
\eeq
that has two Fermi arcs on each open surface.
Then, we perform the following rotational transformation to
the effective Hamiltonian to obtain generally-orientated Fermi arcs as
$
H'_{\text{TR}}(\bm{k})=H_{\text{TR}}(\tilde{U}^{-1}\bm{k})
$
with
\beq
\tilde{U}(\phi)=
\small{\left(\begin{array}{ccc}\frac{\cos\phi}{\sqrt{2}}
&\frac{1}{\sqrt{2}}&-\frac{\sin\phi}{\sqrt{2}}
\\-\frac{\cos\phi}{\sqrt{2}}&\frac{1}{\sqrt{2}}
&\frac{\sin\phi}{\sqrt{2}}\\\sin\phi
&0&\cos\phi\end{array}\right)}.
\eeq
The reason we apply $\tilde{U}(\phi)$ instead of $U(\varphi)$ [cf.~Eq.~\eqref{eq4}] in the $\mathcal{T}$-symmetric case is because of different original positions of the Weyl points in the Brillouin zone to the $\mathcal{P}$-symmetric case.

\begin{figure}
\center
\includegraphics[width=\linewidth]{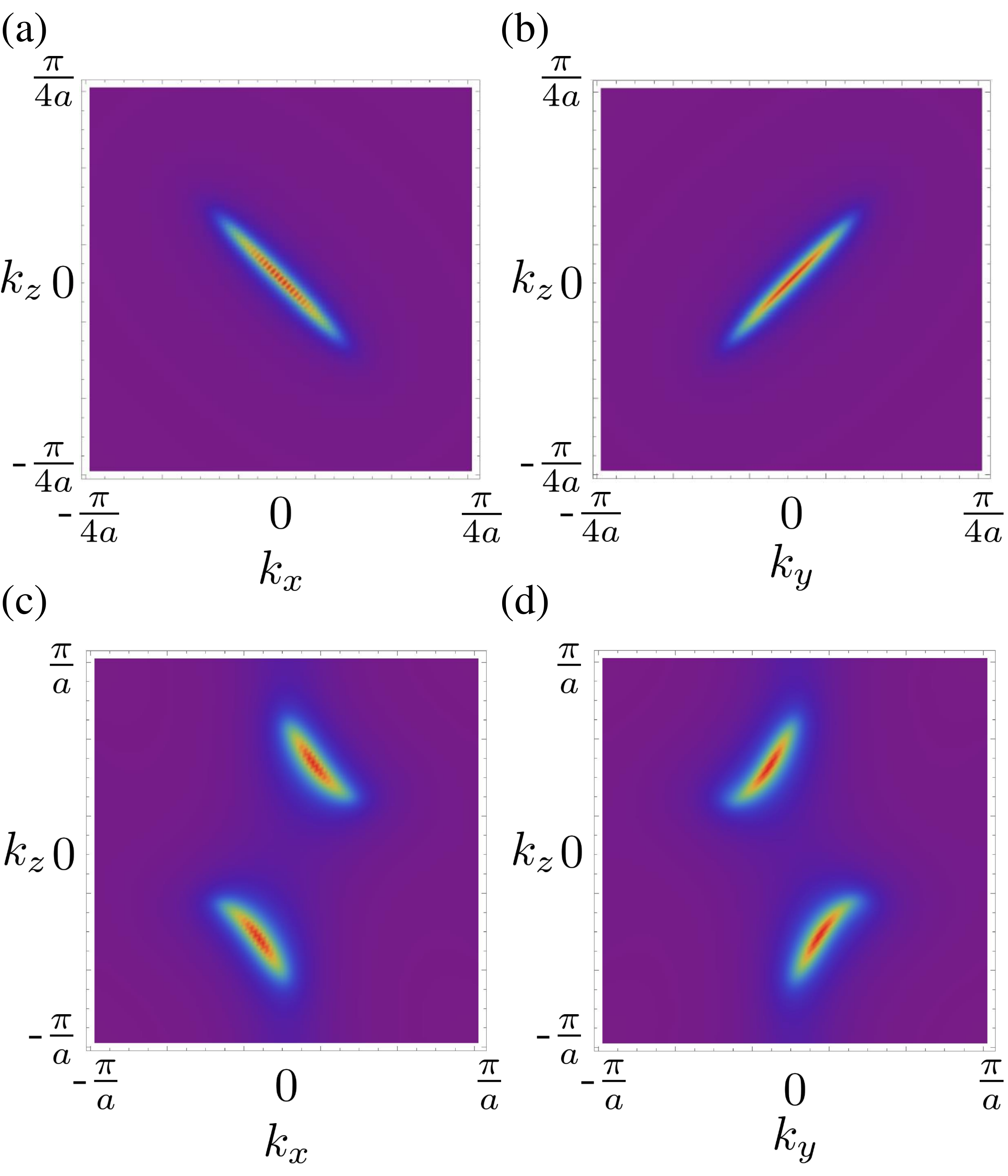}
\caption{Fermi arcs on surfaces I (a) and II (b) of $H'(\bm{k})$ and that on surfaces I (c) and II (d) of $H'_{\text{TR}}(\bm{k})$, with parameters $k_0=0.1$nm$^{-1}$, $M=-1.25$eV$\cdot$nm$^2$ and $\varphi=\cos^{-1}\frac{1}{\sqrt{3}}$ in $H'(\bm{k})$ and $k_0=0.2$nm$^{-1}$, $k_1=\sqrt{2}k_0$, $M=1.25$eV$\cdot$nm$^2$, $\phi=\frac{\pi}{4}$ in $H'_{\text{TR}}(\bm{k})$.}
\label{lattice_Fermi_arcs}
\end{figure}

In the long-wavelength limit, the matching lattice model used in the numerical simulation can be constructed from the effective Hamiltonian through the substitution
$
k_{i=x,y,z}\to a^{-1}\sin k_ia,\quad k_i^2\to2 a^{-2}(1-\cos k_ia),
$
where $a$ is the lattice constant. 
The Fermi arcs of the $\mathcal{P}$-symmetric Weyl semimetal with $\varphi=\cos^{-1}\frac{1}{\sqrt{3}}$ (s.t. $\theta=\frac{\pi}{4}$) and $\mathcal{T}$-symmetric Weyl semimetal with $\phi=\frac{\pi}{4}$ are shown in Fig.~\ref{lattice_Fermi_arcs}.

\begin{figure}
\center
\includegraphics[width=\linewidth]{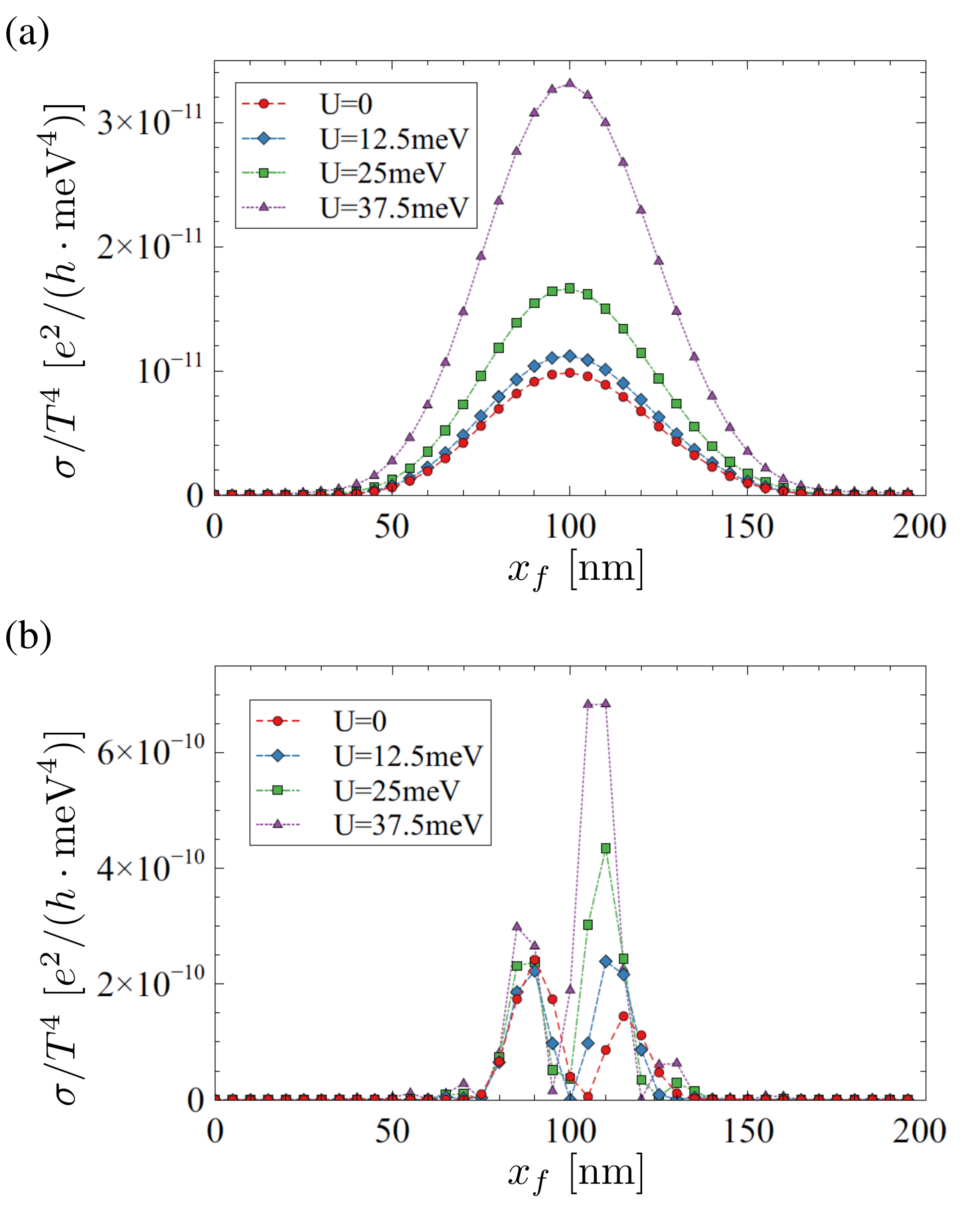}
\caption{Numerical simulations of non-local conductance for (a) $\mathcal{P}$-symmetric Weyl semimetal $H'(\bm{k})$ with different dispersions $U_{\text{I}}=-U_{\text{II}}=U$ and (b) $\mathcal{T}$-symmetric Weyl semimetal $H'_{\text{TR}}(\bm{k})$ with different dispersions $U_{\text{I}}=U_{\text{II}}=-U$. The setup is the same as the one shown in Fig.\ref{fig2}(a), with the scattering region being a Weyl nanowire with a cross-section of $40\times40$ sites with lattice constant $a=5$nm. The conductance is normalized by $T^4$, with $T$ being the tunneling between the leads and the nanowire [cf. Eq.\eqref{T}]. In realistic cases, $T$ is determined by the hopping between the leads and the system as well as the DOS of the system. Parameters are $v=10^6$m/s, $x_i=100$nm, $\varepsilon=0$ and all other parameters the same as Fig.~\ref{lattice_Fermi_arcs}. }
\label{numerics}
\end{figure}

\section{Green's function calculation}\label{gf}
For a fixed energy $E$, the normalized
eigenstate of the surface Hamiltonian is
\begin{equation} \label{eigenstate}
\begin{split}
\psi_{k_z,E}(x,z)=\frac{e^{ik_zz}}{\sqrt{S}}\Big[\theta(-x)e^{ik_xx}
+\theta(x)e^{ik_{x2}x}\Big],
\end{split}
\end{equation}
where the momentum $k_z$ is conserved during the transmission,
and $k_x$ and $k_{x2}$ are solved by $H'_{\text{I}}(k_x,k_z)=E$
and $H'_{\text{II}}(k_{x2},k_z)=E$, respectively.
The function $\theta(\pm x)$ is the
the Heaviside step function defining the two sides of the junction
and $S$ is the combined area of
surfaces I and II.
Without coupling to the terminals, the bare Green's functions can be
constructed as
\beq \label{eqB2}
g^{R,A}_\varepsilon(\bm{r}_f,\bm{r}_i)
&=&\sum_{E}\sum_{k_z}\frac{\psi_{k_z,E}(\bm{r}_f)\psi_{k_z,E}^*(\bm{r}_i)}{\varepsilon-E\pm i0},
\eeq
thus, describing electron propagation from $\bm{r}_i$ to $\bm{r}_f$.
Since the surface states are unidirectional,
we have $g_\varepsilon^R(\bm{r}_i,\bm{r}_f)=0$.
By writing the sums as integrals we obtain Eq. \eqref{g}
in the main text.

The coupling to the $\alpha$ terminal introduces
finite self-energy
to the Green's function of the surface state as
\beq \label{self-energy_inversion}
\Sigma^R_\alpha(\bm{r}_1,\bm{r}_2,\varepsilon)=-i\pi\rho_\alpha(\varepsilon)
|T_\alpha|^2\delta(\bm{r}_1-\bm{r}_\alpha)\delta(\bm{r}_2-\bm{r}_{\alpha}),\nonumber\\
\eeq
The full Green's function can be calculated by Dyson's equation,
which yields Eq. \eqref{G} in the main text.
The linewidth function is defined by $\Gamma_\alpha(\bm{r}_1-\bm{r}_\alpha)=2i \Sigma^R_\alpha(\bm{r}_1-\bm{r}_\alpha)$,
corresponding to Eq. \eqref{L} in the main text.

\section{Numerical simulations of non-local transport experiment}\label{num}

We compare our semiclassical analytical calculation approach with numerical simulations of the non-local transport experiment for both $\mathcal{P}$- and $\mathcal{T}$-symmetric Weyl semimetals $H'(\bm{k})$ and $H'_{\text{TR}}(\bm{k})$ using the numerical package KWANT \cite{groth2014kwant}. To be realistic,
we adopt onsite potential $U_\text{I}\ (U_{\text{II}})$
to the first layer of the lattice
on surface I (II) to introduce dispersion effects,
which results in curved Fermi arcs.
The non-local conductance under different choices of the onsite potentials $U_{\text{I},\text{II}}$ is shown in Fig.~\ref{numerics}. In both cases, the conductance peak value increases with the onsite potential, which is due to the increase of the surface DOS. In addition, in the $\mathcal{T}-$symmetric case, the position of the peak varies for different onsite potential, which is due to the shift of the phase term in $f'_\varepsilon(\bm{r}_f,\bm{r}_i)$ for dispersive Fermi arcs. In both cases, the peak structure persists for dispersive Fermi arcs, in agreement with the analytical results in Figs.~\ref{fig2} and \ref{TR_conductance_FP}.

\section{Introduction of surface dispersion from on-site potential}\label{appen_E}
We show explicitly the on-site potentials $U_\text{I}(U_{\text{II}})$ we adopt on the surface of the Weyl semimetal in Appendix \ref{num} result in the dispersion terms in Eq.\eqref{HS}. Note that the surface states on surface I possess a $\mathbf{k}$-dependent wave function $\psi_{k_x,k_z}(y)$ that has some spatial profile along $y$ direction. Under surface potential $U_\text{I}(y)$, the potential that the states feel can be evaluated by the overlap integral ${{U}_{\text{I}}}(k)=\mathop{\int }_{0 }^{\infty}{{U}_{\text{I}}}(y)|{{\psi }_{k_x,k_z}}(y){{|}^{2}}dy$, which is $\mathbf{k}$-dependent and thus serves as an effective dispersion ${{\varepsilon}_{\text{I}}}(k_x,k_z)={{U}_{\text{I}}}(k_x,k_z)$. In our model, the surface states' wavefunctions satisfy $|\psi_{k_x,k_z}(y)|=|\psi_{-k_x,-k_z}(y)|$ so that the effective dispersion is an even function of $k_x$ and $k_z$, and takes the form $\varepsilon_y(k_x,k_z)=\varepsilon_0-d(k_x^2+k_z^2)$ to the second order in $k_x$ and $k_z$. In addition, the dispersion term should vanish at Weyl points where the surface states spread in the whole bulk, which leads to $\varepsilon_0=d(1-\sin^2\varphi/2)k_0^2$. Therefore the surface potential $U_{\text{I}}$ leads to the dispersion $\varepsilon_x$ in Eq.\eqref{HS}. Similarly, the surface potential $U_{\text{II}}$ leads to the dispersion $\varepsilon_y$ in Eq.\eqref{HS}

\end{appendix}

\bibliographystyle{apsrev4-1-etal-title}

\bibliographystyle{apsrev4-1}

\end{document}